\documentclass[twocolumn,showpacs,prl]{revtex4}
\usepackage{graphicx}
\usepackage{natbib}

\begin{document}
\newcommand{\Fig}{Fig.~}
\newcommand{\Figs}{Figs.~}
\newcommand{\Figfirst}{Figure~}
\newcommand{\Rb}{${}^{87}$Rb }
\newcommand{\fxm}[1]{{\bf !---!FIXME #1 !---!}}
\newcommand{\ds}{}
\newcommand{\etal}{\textit{et al.~}}
\newcommand{\correction}[2]{\fbox{\fbox{#1} {#2}}}

\author{Eugeniy E.\ Mikhailov}
    \email{evmik@tamu.edu}
\author{Vladimir A.\ Sautenkov}
\author{Yuri V.\ Rostovtsev}
\author{George R.\ Welch}

\affiliation{
        Department of Physics and Institute of Quantum Studies,
        Texas A\&M University,
        College Station, Texas 77843-4242
}

\title{ 
	Absorption resonance and large negative delay in Rb vapor 
	with buffer gas.
} 

\begin{abstract} 
We observe a narrow, isolated, two-photon absorption resonance
in \Rb for large one-photon detuning in the presence of
a buffer gas.  In the absence of buffer gas, a standard
$\Lambda$ configuration of two laser frequencies gives rise
to electromagnetically induced transparency (EIT) for all
values of one-photon detuning throughout the inhomogeneously
(Doppler) broadened line.  However, when a buffer gas is added
and the one-photon detuning is comparable to or greater than
the Doppler width, an absorption resonance appears instead
of the usual EIT resonance.  We also observe large negative
group delay ($\approx -300~\mu\mathrm{s}$ for a Gaussian pulse
propagating through the media with respect to a reference pulse
not affected by the media), corresponding to a superluminal
group velocity $v_g= -c/(3.6 \times 10^6)=-84~\mathrm{m/s}$.
\end{abstract} 

\date{\today}
\maketitle



	Quantum coherence effects have attracted great attention
the last few years.  Phenomena such as ultra-low optical
group velocity~\cite{hau99, kash99, budker99, godone02pra},
super-luminal group velocity~\cite{godone02pra,wang2000nature,
akulshin03pra}, enhanced nonlinear optical
effects~\cite{harris99prl}, and quantum information
storage~\cite{phillips01prl, hau01nature, zibrov02prl} have
all been studied under the condition of electromagnetically
induced transparency (EIT)~\cite{harris'97pt}.  More recently,
complementary coherence effects such as electromagnetically
induced absorption (EIA) have been predicted and
studied~\cite{akulshin03pra,akulshin'98, taichenachev'00pra,
ye'02}.
%
%
In the Letter, we present the experimental observation of a new
narrow absorption resonance appearing at large optical density
and large detuning in Doppler broadened media with buffer gas.

	In typical experiments, the narrow transmission
linewidth in EIT is limited by the relaxation rate of the
ground-state coherence, which is usually determined by the
interaction time of an atom with the applied laser radiation.
Two common methods for increasing the interaction time are by
use of wall coatings~\cite{budker'98}
%
%
that allow an atom to maintain its coherence while it travels
into and out of the interaction region many times, and by use
of a buffer gas that allows the atom to diffuse out of the
interaction region slowly by velocity changing collisions that
still preserve the ground-state coherence~\cite{brandt'97,
helm'01}.
%

	For a $\Lambda$-type EIT system, the dependence
of the EIT resonance on one-photon laser detuning has
been studied experimentally~\cite{wynands2003shapes} and
theoretically~\cite{taichen2002}.
%
%
They show that in the limit of high buffer gas pressure,
when the decay rate of the upper level is comparable with the
Doppler broadening, the EIT resonance shape can be described
by a Lorentzian plus a dispersion-like curve:
\begin{equation}
    \label{res_shape_fit}
        f(\delta)=\gamma \frac{A\gamma + B (\delta-\delta_0)}
	    {\gamma^2 + (\delta-\delta_0)^2} + C
\end{equation}
where $\gamma$ is the width of the resonance, $\delta$ is
the two photon detuning, $\delta_0$ is one-photon dependent
resonance shift with respect to resonance position for
zero one-photon detuning, $A$ is the amplitude of the
Lorentzian part, $B$ is the amplitude of the dispersion-like
part, and $C$ is an offset.  However previous experimental
studies~\cite{wynands2003shapes} have only seen the resonance
shape become somewhat assymetric while still maintaining an
EIT resonance-like shape, in other words $A>0$ and $|B/A|<2$.

	In our experiment we achieve an absorption-like
resonance ($A<0$) by detuning the drive laser into the
wings of the Doppler distribution in a cell with buffer gas.
For our conditions, no absorption is found without buffer gas.
The absorption resonance is accompanied by large anomalous
dispersion so that negative group delay is observed for a pulse
propagating through the medium.  It is important to stress that
the previous observation of EIA in Refs.~\cite{akulshin03pra,
akulshin'98, taichenachev'00pra, ye'02} and the enhanced
absorption seen in the Hanle effect~\cite{dancheva'00}
%
%
have a different nature than what we observe here, since
those previous observations require that the degeneracy of
the ground state be lower than that of the excited state~\cite{taichenachev'00pra}, (i.e. $F < F'$) for a drive field transition, 
which is not necessary in our experiments.



\begin{figure} 
\includegraphics[angle=0, width=0.70\columnwidth]
	{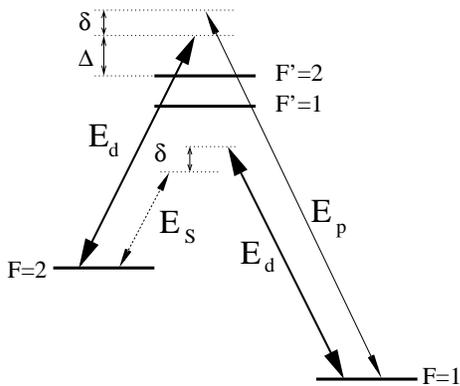}
\caption{
	Three-level interaction scheme of three laser fields
	with \Rb atoms.  The long-lived coherence is created
	on the hyperfine split ground-state sublevels with
	a strong driving field $\mathrm{E_{d}}$ and a weak
	probe field $\mathrm{E_{p}}$.  A weak Stokes field
	$\mathrm{E_{s}}$ is also present as a by-product of
	the generation of the probe field.  $\Delta$ is the
	one-photon detuning of the drive and probe laser from
	their respective atomic transitions, and $\delta$
	is the two-photon detuning, which is scanned.
	\label{levels.fig}
}
\end{figure}

	We use a weak probe and a strong drive (or coupling)
field in a $\Lambda$ configuration of the two ground-state
levels $5S_{1/2}F=1,2$ an the excited $5P_{1/2}F=2$ state of
\Rb as shown in \Fig\ref{levels.fig}.  
We operate in the power-broadened regime
where $\Omega_d > \sqrt{\gamma\gamma_{bc}}$ where $\Omega_d$ is the
drive laser Rabi frequency, $\gamma$ is the radiative decay
rate as defined above, and $\gamma_{bc}$ is the decay rate of
ground-state coherence.  The resulting ground-state coherence
gives rise to a narrow EIT resonance of the probe field in the
vicinity of two-photon resonance ($\delta=0$).  This coherence
still exists even for one-photon detunings ($\Delta$) comparable
to or somewhat greater than the inhomogeneous Doppler width
of the medium.

	Because of the way our probe field is generated
(discussed below) a second $\Lambda$ configuration
consisting of a weak Stokes component and the drive field
(see \Fig\ref{levels.fig}) also is present.  However, this
system is detuned far from resonance, and does not affect our
system notably.  (Nonetheless, the propagation properties
of the Stokes component are rather interesting and will be
discussed elsewhere.)



\begin{figure}[b] 
\includegraphics[angle=0, width=1.00\columnwidth]
	{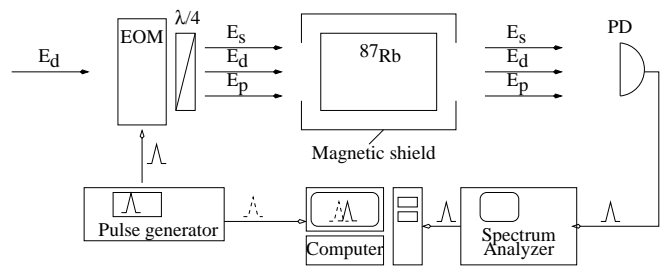}
\caption{
	Schematic of the experimental setup.
	\label{simple_setup}
}
\end{figure}

	Our experimental setup is shown schematically in
\Fig\ref{simple_setup}.  We use an external cavity diode laser
tuned to the vicinity of the $5S_{1/2} \rightarrow 5P_{1/2}$
($D_1$) transition of \Rb.  The laser is referred to as the
driving field ($E_{d}$) in \Fig\ref{levels.fig}.  
Detuning of the drive laser($\Delta$) changes from zero up to 2~GHz and is 
always positive (for any $\Delta > 0$ detuning from upper levels $F'=1,2$ 
is larger then for zero detuning as it shown in \Fig\ref{levels.fig}).
Sidebands are
generated on the drive laser by an electro-optic modulator
(EOM) which is tunable in the vicinity of the 6.835~GHz
ground state splitting.  The drive laser is tuned to the
$F=2\rightarrow F'=2$ transition so that the upper sideband
serves as the probe field, and is tunable in the vicinity of the
$F=1\rightarrow F'=2$ transition.  The lower sideband (Stokes
component) is far off resonance.  After the EOM, all fields pass
through a single-mode optical fiber to obtain a clean Gaussian
spatial mode.  The laser is collimated to a diameter of 7 mm,
and circularly polarized with a $\lambda/4$ wave-plate right
before the cell.  The cell itself is surrounded by 3 layer
magnetic shield which suppresses the laboratory magnetic field.
The cell is heated to $60-70 {}^oC$ to control the density of
\Rb vapor.

	Before the EOM, part of the drive laser is split
from the main beam, shifted in frequency by a small amount
(60~MHz) with an acousto-optic modulator and deflected around
the cell.  This shifted beam is recombined with the light
transmitted through the cell and all the fields are detected
on a fast photodiode.  This (heterodyne) procedure allows
us to separately detect the transmitted probe and Stokes
fields.~\cite{kash99}.

	We measure the EIT spectrum (transmission as a function
of two-photon detuning $\delta$ for various one photon detunings
of the driving field ($\Delta$).  We also measure the group
delay of a pulse propagating through the cell.  This is done
with a programmable pulse synthesizer, which modulates the
microwave generator for the EOM.  We thus produce a Gaussian
(temporal) pulse in the power of the drive field sidebands (the
probe field).  The delay time is extracted by data taking and
analyzing software on a computer which collects the separate
probe and reference signals.

%
%
\begin{figure}[t] 
\includegraphics[angle=0, width=1.00\columnwidth]
{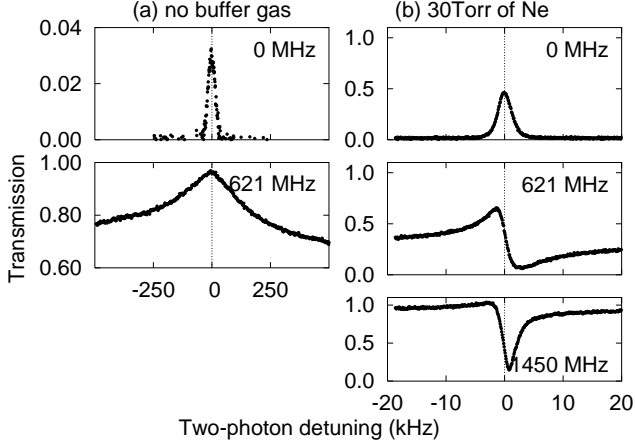}
\caption{
	Transmission of probe field as a function of
	two-photon detuning $\delta$ (measured in kHz)
	for various one-photon detunings $\Delta$.  (a) \Rb
	cell with no buffer gas (vacuum).  (b) \Rb cell with
	$30$~Torr of Ne.  The vertical scales are normalized in such a way 
	that 0 correspond to zero transmission and 1 correspond to zero 
	absorption (or total transparency). 
	Experimental parameters:
	(a)
	$T=66.4^o\mathrm{C}$, $N=4.2\times 10^{11}~{\mathrm{cm}}^{-3}$;
	Total power in to cell $630 \mu\mathrm{W}$, out $40 \mu\mathrm{W}$;
	cell length=4.7~cm.
	(b)
	$T=67.7^o\mathrm{C}$, $N=4.7\times 10^{11}~{\mathrm{cm}}^{-3}$;
	Total power into cell $\approx 400 \mu\mathrm{W}$;
	cell length=2.5~cm.
	\label{eit_shape_example}
}
\end{figure} 


	We first measure various transmission spectra for
a \Rb cell with no buffer gas.  Our measurements, shown in
\Fig\ref{eit_shape_example}a show that the EIT resonance
maintains its transmission-like shape as the drive and probe
are detuned from one-photon resonance.  When the lasers are
far from resonance (bottom trace in \Fig\ref{eit_shape_example}a),
the lasers interact with a very small number of atoms, and the
width of the EIT resonance is determined by power broadening
by the drive laser ($\mathrm{width}\approx\Omega^2/\gamma$
where $\Omega$ is the drive Rabi frequency and $\gamma$ is
the Doppler width).  When the lasers are on resonance, the
effective optical density is much higher, and the EIT width
is much lower than the power-broadened width~\cite{lukin97prl,
sautenkov99las,javan'02}.

	Next we replace the vacuum cell by a cell
with 30~Torr of Ne buffer gas.  Results are shown in
\Fig\ref{eit_shape_example}b.  We find that the resonance
spectra for large one-photon detuning ($\Delta$) are
changed dramatically from the vacuum case just described.
We see that as the one-photon detuning is increased, the EIT
resonance passes through a dispersion-like shape and into an
absorption-like shape.

\begin{figure} 
\includegraphics[angle=0, width=1.00\columnwidth]
	{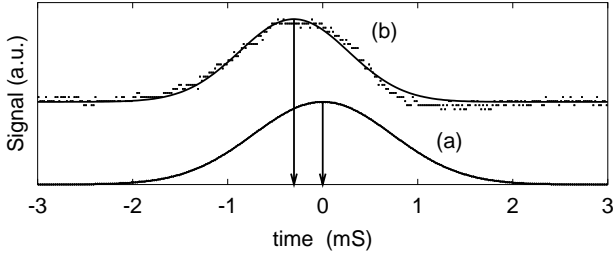} 
\caption{
	Demonstration of negative pulse delayed time.  (a) The
	probe field is given a Gaussian temporal pulse shape.
	(b) The transmitted probe field, showing negative time
	delay before transmission.  Experimental conditions
	same as for \Fig\ref{eit_shape_example}b.
	\label{pulse_delay}
}
\end{figure}

	Second, we find that this absorption-like resonance
is accompanied by steep anomalous dispersion which results
in a large negative propagation delay of the probe pulse
with respect to a reference pulse that is not delayed by the
medium.  Sample pulses are shown in \Fig\ref{pulse_delay}.
These data show a $-300~\mu\mathrm{s}$ delay produced by a
$2.5~\mathrm{cm}$ long cell, implying a superluminal group
velocity $v_g=-84~\mathrm{m/s}$.

\newpage

	In order to gain physical insight into this new
phenomenon we have made numerical simulations based on the
density matrix equations coupled with a Maxwell equation
description of propagation effects in steady state for low
intensity of the probe field.
For numerical simulation we use a 
3-level atomic model coupled with probe and drive 
fields ($\Lambda$ scheme). 

%
The density matrix equations are
the following:
\begin{eqnarray}
\label{rho_eit_eq_bia}
\dot{\rho}_{bb} & = &
    i \Omega_p^* \rho_{ab} - i \Omega_p \rho_{ba}
    + \gamma_r \rho_{aa} - \gamma_{bc} \rho_{bb} + \gamma_{bc} \rho_{cc},
        \\
\dot{\rho}_{cc} & = &
    i \Omega_d^* \rho_{ac} - i \Omega_{d} \rho_{ca}
    + \gamma_r \rho_{aa} - \gamma_{bc} \rho_{cc} + \gamma_{bc} \rho_{bb},
        \\
\dot{\rho}_{ab} &=& - \Gamma_{ab} \rho_{ab} + i \Omega_p (\rho_{bb} - \rho_{aa} )
        + i \Omega_d \rho_{cb},
        \\
\dot{\rho}_{ca}  &=& - \Gamma_{ca} \rho_{ca} + i \Omega_d^* (\rho_{aa} - \rho_{cc} )
        - i \Omega_p^* \rho_{cb},
        \\
\dot{\rho}_{cb} &=& - \Gamma_{cb} \rho_{cb} - i \Omega_p \rho_{ca}
        + i \Omega_d^* \rho_{ab},
    \label{rho_lambda_last_eq_bia}
\end{eqnarray}
where $\Omega_d=\wp_{ac}\mathrm{E_{p}}/\hbar$ and
$\Omega_p=\wp_{ab}\mathrm{E_{d}}/\hbar$ are the Rabi frequencies
of the drive and probe fields. The generalized decay rates are
defined as:
\begin{eqnarray}
\Gamma_{ab}&=&\gamma+i(\Delta+\delta), \\
\Gamma_{ac}&=&\gamma+i\Delta,  \\
\Gamma_{cb}&=&\gamma_{bc}+i\delta.
\end{eqnarray}
Here $\gamma = \gamma_r+\gamma_{p}$ is the polarization decay
rate, $\gamma_r$ is the radiative decay rate of the excited state,
$\gamma_{p}$ is dephasing  rate, 
and $\gamma_{bc}$ is the inverse lifetime of the coherence between
ground states $|b\rangle$ and $|c\rangle$. Here we recall
that the presence of the buffer gas affects values of both
$\gamma_{bc}$ and $\gamma$.  
On one hand, as mentioned
above, it allows preservation of the ground-state coherence longer,
but on the other hand it broadens the optical transition, since
$\gamma_{p}$ grows linearly with buffer gas
pressure~\cite{vanier_book}.

Solving these equations in the steady state regime and assuming
$|\Omega_p| \ll |\Omega_d|$, we obtain the following expression for
the linear susceptibility of the probe field:


\begin{equation}
\label{chi1}
\chi_{ab}(\Delta)=i\eta
\frac{\Gamma_{cb}(\rho^{(0)}_{aa}-\rho^{(0)}_{bb})+\displaystyle{\frac{|\Omega_d|^2}{\Gamma_{ca} }}(\rho^{(0)}_{aa}-\rho^{(0)}_{cc})}
{\Gamma_{ab}\Gamma_{cb}+|\Omega_d|^2},
\end{equation}
where $\eta = \frac{3}{8\pi}N\lambda^2\gamma_r$, $N$ is the
\Rb density, $\lambda$ is the wavelength of the probe field,
and $\rho_{aa}^{(0)}$, $\rho_{bb}^{(0)}$, $\rho_{cc}^{(0)}$
are the solution of above equations in case when $\Omega_p=0$.  
Then to take into account propagation and Doppler averaging, 
the Maxwell equation is given by 
\begin{equation}
\ds{\partial\Omega_p\over\partial z} = -i\Omega_p\int\chi_{ab}(\Delta + kv) dv
\end{equation}
has been solved numerically, where $k$ is the wavenumber of either
probe or drive field.

\begin{figure} 
\includegraphics[angle=0, width=1.00\columnwidth]
	{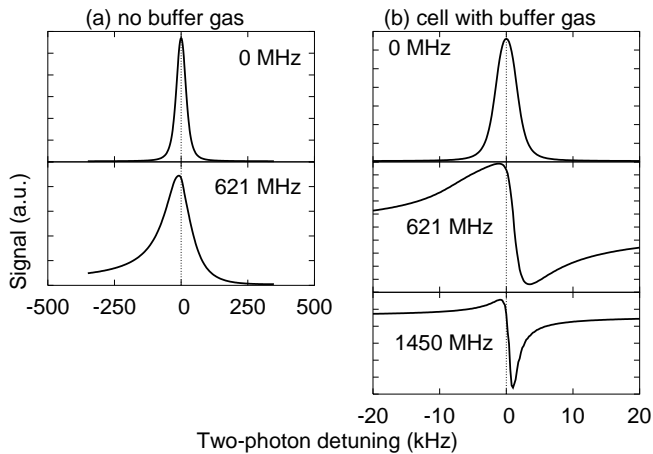}
\caption{
	Calculated transmission of probe field as a function
	of two-photon detuning $\delta$ (measured in kHz)
	for various one-photon detunings $\Delta$.  (a) \Rb
	cell with no buffer gas (vacuum).  (b) \Rb cell with
	$30$~Torr of Ne.  The vertical scales are arbitrary.
	\label{eit_theory}
}
\end{figure}

%
%
We model the effect of the buffer gas by adding an additional
homogeneous broadening for the optical transitions $\gamma_p$
and decreasing of the time-of-flight limit for the for
the hyperfine coherence decay rate $\gamma_{bc}$ based on
the atomic diffusion in the buffer gas.  In our simulation
for EIT in the absence of buffer gas, we use $\gamma_p = 0$
and $\gamma_{bc}/2\pi = 10$~kHz.  For our simulation of the
effects of buffer gas, we use $\gamma_p/2\pi = 120$~MHz and
$\gamma_{bc}/2\pi = 1$~kHz.  The results of these simulations
are shown in \Fig\ref{eit_theory} agree well with the
experimental results of \Fig\ref{eit_shape_example}.  
We model the effect of the buffer gas by adding an additional
homogeneous broadening for the optical transitions $\gamma_p$
using the cross-sections for broadening from~\cite{happer72}
and decreasing of the time-of-flight limit for the for
the hyperfine coherence decay rate $\gamma_{bc}$ based on
the atomic diffusion in the buffer gas~\cite{happer72}.  In our
simulation
for EIT in the absence of buffer gas, we use $\gamma_p = 0$
and $\gamma_{bc}/2\pi = 10$~kHz.    For our simulation of the
effects of buffer gas, we use $\gamma_p/2\pi = 120$~MHz and
$\gamma_{bc}/2\pi = 1$~kHz~\cite{happer72}.
The results of these simulations
are shown in \Fig\ref{eit_theory} agree well with the
experimental results of \Fig\ref{eit_shape_example}.
	

	Previous measurements involving one-photon detuning in
EIT~\cite{wynands2003shapes} differ from ours in that low laser
power was used on the $D_2$ line of Cesium, which includes
a closed cycling transition, and also upper-state hyperfine
structure that is covered by the Doppler width.  Our numerical
model shows the absence of the absorption resonance 
when additional homogeneous broadening parameter ($\gamma_p$) is small 
and
at low
probe and drive intensity.  This may be part of the reason
why the absorption resonance was not observed in that work.
Due to limited sensitivity of our setup we are not able to check
this experimentally.  The lowest total power for which we have
data is 50~$\mu$W, which is still in the power-broadened regime.



	In summary, we have observed a new, isolated, narrow,
two-photon absorption resonance that appears under EIT
conditions when a buffer gas is used to narrow the two-photon
resonance line and the one-photon detuning is comparable
to or greater than the inhomogeneous Doppler linewidth.
The effect occurs when homogeneous broadening 
(due to collisions with buffer gas) is of the order of inhomogeneous 
(Doppler) broadening, and the effect 
does not occur in the absence of buffer gas at room temperature. 
Although, let us note here that for cold atoms 
the Doppler broadening can also be of the order 
of or even much smaller than the homogeneous broadening, 
and, in this case, similar 
absorption resonances should be observable. This technique 
may provide a new tool to study BEC. 
Unlike previous observations of Electromagnetically Induced
Absorption, this resonance does not require that the degeneracy
of the ground state be lower than that of the excited state.
The absorption resonance reported here is accompanied by steep
anomalous dispersion giving rise to very large negative group
delay. 



	We thank A.\ B.\ Matsko, Irina Novikova, A.\ M.\
Akulshin, and M.\ O.\ Scully for useful and stimulating
discussions.  This work was supported by the Office of Naval
Research.


\bibliographystyle{apsrev}


\end{document}